\documentclass[amsmath, amssymb,10pt,aps,prl,reprint,notitlepage,showpacs,superscriptaddress,longbibliography]{revtex4-1}
\usepackage{graphicx}
\usepackage{calc}
\usepackage{braket}
\usepackage{ulem}   
\usepackage{slashed}
\usepackage{wasysym}
\usepackage{siunitx}
\usepackage{dsfont}
\usepackage{amsthm,amsmath,amsfonts,amssymb,verbatim,xcolor}
\usepackage{graphicx}
\usepackage{subfigure}
\usepackage{bm}
\usepackage{epsfig,slashed}
\usepackage[T1]{fontenc}
\usepackage{ulem}
\usepackage{soul} 
\usepackage[colorlinks=true,citecolor=blue,linkcolor=blue,urlcolor=blue]{hyperref}
\newcommand{\tr}{\mathop{\mathrm{tr}}}

\newcommand{\bq}{{\bf q}}

\newcommand{\bd}{{\bf d}}

\newcommand{\bR}{{\bf R}}
\newcommand{\bS}{{\bf S}}

\newcommand{\bM}{{\bf M}}
\newcommand{\bQ}{{\bf Q}}

\newcommand{\be}{\begin{equation}}
\newcommand{\ee}{\end{equation}}
\newcommand{\beg}{\begin{gather}}
\newcommand{\eeg}{\end{gather}}
\newcommand{\beq}{\begin{eqnarray}}
\newcommand{\eeq}{\end{eqnarray}}
\newcommand{\bea}{\begin{align}}
\newcommand{\eea}{\end{align}}
\newcommand{\beqq}{\begin{eqnarray*}}
\newcommand{\eeqq}{\end{eqnarray*}}

\begin{document}

\setstcolor{red} 

\title{Signatures of Z$_3$ Vestigial Potts-nematic order in van der Waals antiferromagnets }

\author{Zhuoliang Ni}
\affiliation{Department of Physics and Astronomy, University of Pennsylvania, Philadelphia, Pennsylvania 19104, USA}%
\author{Daniil S. Antonenko}
\affiliation{Department of Physics, Drexel University, Philadelphia, Pennsylvania 19104, USA}
\author{W. Joe Meese}
\affiliation{School of Physics and Astronomy, University of Minnesota, Minneapolis, Minnesota 55455, USA}%
\author{Qi Tian}
\affiliation{Department of Physics and Astronomy, University of Pennsylvania, Philadelphia, Pennsylvania 19104, USA}%
\author{Nan Huang}
\affiliation{Department of Materials Science and Engineering, University of Tennessee, Knoxville, TN 37996, U.S.A.}
\author{Amanda V. Haglund}
\affiliation{Department of Materials Science and Engineering, University of Tennessee, Knoxville, TN 37996, U.S.A.}
\author{Matthew Cothrine}
\affiliation{Department of Materials Science and Engineering, University of Tennessee, Knoxville, TN 37996, U.S.A.}
\author{David G. Mandrus}
\affiliation{Department of Materials Science and Engineering, University of Tennessee, Knoxville, TN 37996, U.S.A.}
\author{Rafael M. Fernandes}
\affiliation{School of Physics and Astronomy, University of Minnesota, Minneapolis, Minnesota 55455, USA}%
\author{Jörn W. F. Venderbos}
\affiliation{Department of Physics, Drexel University, Philadelphia, Pennsylvania 19104, USA}
\affiliation{Department of Materials Science \& Engineering, Drexel University, Philadelphia, PA 19104, USA}%
\author{Liang Wu}
\affiliation{Department of Physics and Astronomy, University of Pennsylvania, Philadelphia, Pennsylvania 19104, USA}%

\date{\today}

\begin{abstract}

Layered van der Waals magnets have attracted much recent attention as a promising and versatile platform for exploring intrinsic two-dimensional magnetism. Within this broader class, the transition metal phosphorous trichalcogenides $M$P$X_3$ stand out as particularly interesting, as they provide a realization of honeycomb lattice magnetism and are known to display a variety of magnetic ordering phenomena as well as superconductivity under pressure. One example, found in a number of different materials, is commensurate single-$Q$ zigzag antiferromagnetic order, which spontaneously breaks the spatial threefold $(C_3)$ rotation symmetry of the honeycomb lattice. The breaking of multiple distinct symmetries in the magnetic phase suggests the possibility of a sequence of distinct transitions as a function of temperature, and a resulting intermediate $\mathbb{Z}_3$-nematic phase which exists as a paramagnetic vestige of zigzag magnetic order -- a scenario known as vestigial ordering. Here, we report the observation of key signatures of vestigial Potts-nematic order in rhombohedral FePSe$_3$. By performing linear dichroism imaging measurements---an ideal probe of rotational symmetry breaking---we find that the $C_3$ symmetry is already broken above the N\'eel temperature. We show that these observations are explained by a general Ginzburg-Landau model of vestigial nematic order driven by magnetic fluctuations and coupled to residual strain. An analysis of the domain structure as temperature is lowered and a comparison with zigzag-ordered monoclinic FePS$_3$ reveals a broader applicability of the Ginzburg-Landau model in the presence of external strain, and firmly establishes the $M$P$X_3$ magnets as a new experimental venue for studying the interplay between Potts-nematicity, magnetism and superconductivity.
\end{abstract}

\maketitle

Vestigial order refers to an unconventional and rare type of ordering that can arise when a primary order parameter remains fluctuating and thus disordered, but a secondary order parameter---a composite of the primary order parameter---condenses \cite{Nie2014,Eduardo2015,Fernandes2019}. While various ordered phases have been proposed to support vestigial order \cite{Nie2017,Demler2017,Sandvik2020,Orth2022,Scheurer2023}, a prime example of this phenomenon can be found in magnetically ordered systems that, in addition to the fundamental symmetries associated with magnetism (i.e., spin rotation and time-reversal symmetry), also break symmetries of the crystal lattice, such as crystal rotation symmetry. Such a pattern of symmetry breaking does not occur in simple conventional N\'eel antiferromagnets, but instead requires a special kind of magnetic order described by a multicomponent magnetic order parameter \cite{Fernandes2019}. The emergence of such order typically requires some form of magnetic frustration \cite{Chandra1990}. Multicomponent magnets allow for distinct types of magnetic fluctuations, for instance in the nematic channel, and the condensation of symmetry-breaking fluctuations can lead to long-range vestigial order.
The high-temperature iron-pnictide superconductors, in which a transition from tetragonal to orthorhombic symmetry is observed above the spin density wave ordering temperature, provide a well-known experimental realization of broken rotation symmetry driven by magnetic fluctuations \cite{Chu2010,Fernandes2014,Bohmer2022}. The resulting orthorhombic paramagnetic phase is characterized by a spontaneously broken Ising symmetry and thus realizes a vestigial Ising-nematic phase \cite{Fang2008,Xu2008,Pnictides_PRB}. 

A qualitatively different type of vestigial nematic order can arise in hexagonal antiferromagnets that spontaneously break threefold crystal rotation symmetry \cite{Little2020}. Rather than an Ising symmetry, threefold rotations realize a $\mathbb{Z}_3$ symmetry, and as a result, nematic order in systems with hexagonal symmetry belongs to the three-state Potts universality class \cite{Fernandes2019}. Compared to the more familiar Ising case, the study of Potts-nematicity is still in its infancy, spurred by a remarkable surge of reports of threefold rotation symmetry breaking in a variety of correlated systems, from magnets \cite{Little2020} to cold atoms \cite{Jin2021} to superconductors \cite{Cao2021}.  Moreover, superconductivity has been reported in  hexagonal antiferromagnets \cite{wang2018emergent}, and therefore these systems provide a rich platform to study the intertwining between magnetism, superconductivity and vestigial order. 
 
\begin{figure}
\centering
\includegraphics[width=0.5\textwidth]{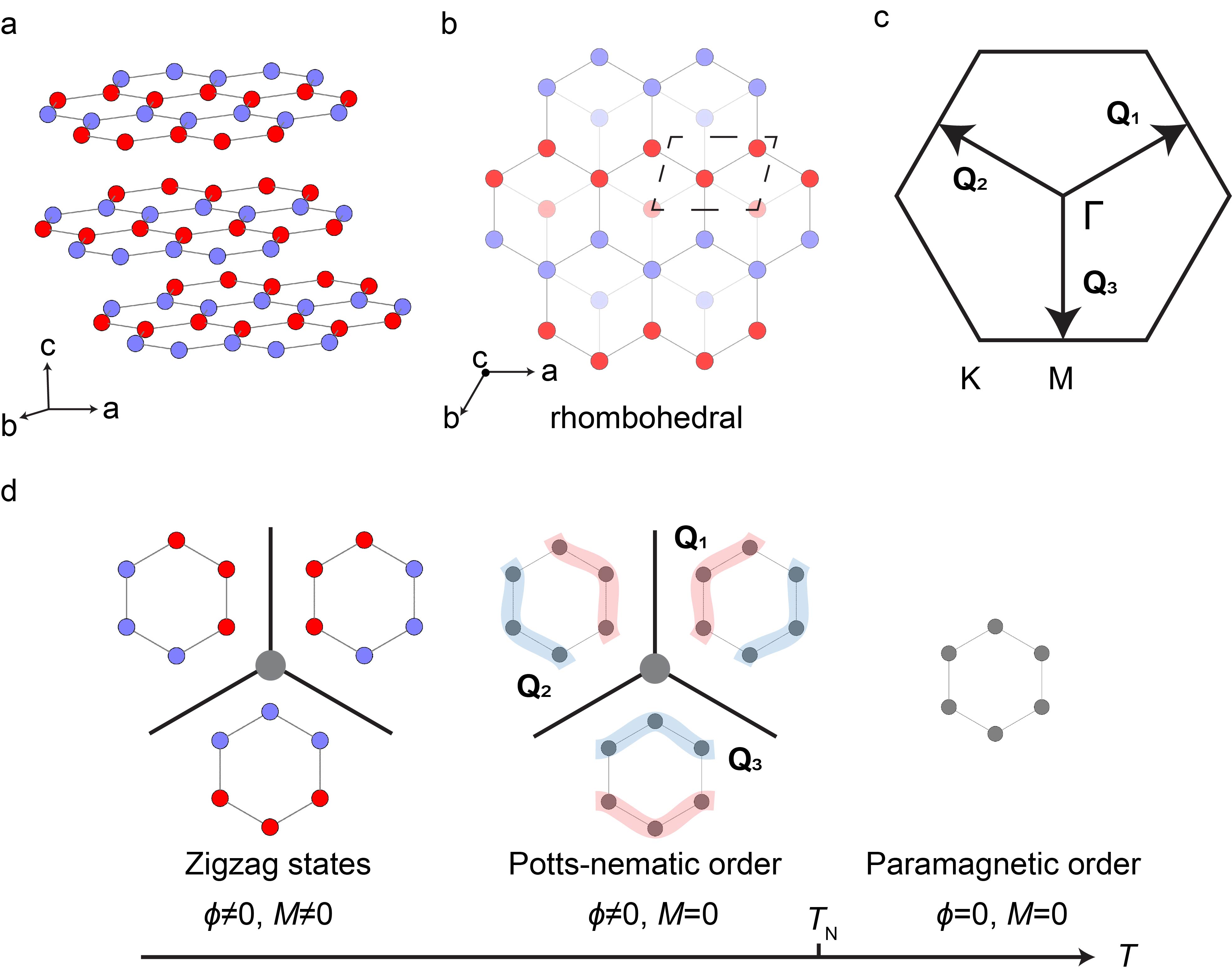}
\caption{ (a) Side view of the lattice and magnetic structure (zigzag order) of the Fe atoms in FePSe$_3$. The red and violet colors represent two opposite spin orientations. (b) Top view of the lattice and magnetic structure of the Fe atoms in FePSe$_3$. The more transparent atoms are those on the lower layer. The formation of zigzag order below $T_N$ breaks the C$_3$ symmetry of the lattice. The dashed box represents an unit cell of the crystal lattice. (c) Top view of the one-layer Bravais lattice of FePSe$_3$. The zigzag order is a single-$Q$ phase with ordering vector at one of the three $M$ points (\textbf{Q}$_3$ is shown in this figure).  (d) Possible evolution of the zigzag and Potts-nematic orders when $T$ increases. The zigzag states and Potts-nematic order develops below $T_{\mathrm{N}}$. }

\label{fig1}
\end{figure}

In this work we reveal experimental signatures of vestigial Potts-nematic order in the van der Waals antiferromagnet FePSe$_3$. The compound FePSe$_3$ belongs to the class of transition metal phosphorous trichalcogenides $M$P$X_3$ (M = Mn, Ni, Fe, Co; X = S, Se)~\cite{joyprb1992,sivadasprb15,chittariprb16}, which form quasi-2D layered structures with weak van der Waals bonding. The $M$P$X_3$ materials are known to exhibit a variety of long-ranged antiferromagnetic orders, and since the magnetic transition metal sites within each layer form a honeycomb lattice, they provide an appealing realization of model honeycomb magnets. As such, they have attracted much attention as a possible platform for exploring intrinsic 2D magnetism~\cite{wiedenmannssc81,kurosawajpsj1983,ruleprb07,lanconprb16,ressoucheprb10,leeaplm2016,wang2dmat16,kimnatcomm19,ninatnano21,chuprl20,niprl2021}. Within the $M$P$X_3$ family, the Fe compounds exhibit collinear ``zigzag''-ordered antiferromagnetic (AFM) phases below the N\'eel temperature ($T_N$), characterized by mutually anti-aligned ferromagnetic zigzag chains, as shown in Fig.\ref{fig1}. A sharp change of the lattice coefficients at $T_N$ in both FePS$_3$ and FePSe$_3$ has been reported, indicating that a lattice phase transition is intertwined with the magnetic transition~\cite{jernbergjmmm84}. Finally, superconductivity develops in the Fe compounds under applied pressure \cite{wang2018emergent}.

The zigzag magnetic orders observed in the Fe compounds are special for two reasons: they suggest strong magnetic frustration caused by competing exchange interactions, and, most importantly, they break the threefold crystalline rotational symmetry $C_3$ of the individual honeycomb layers. From the perspective of a monolayer there are three possible zigzag directions, corresponding to three symmetry-related AFM ordering vectors. As a result, the formation of zigzag order spontaneously breaks $C_3$ rotation symmetry. In bulk form, however, FePS$_3$ and FePSe$_3$ differ in the way that the layers are stacked, giving rise to bulk structures with different symmetries. Whereas FePS$_3$ has monoclinic stacking order characterized by space group $C2/m$~\cite{zhangnanolett21,Xzhangnanolett21}, FePSe$_3$ has rhombohedral $ABC$ stacking with space group $R\overline{3}$, as shown in Fig.\ref{fig1}(b). The absence of $C_3$ symmetry in bulk FePS$_3$ implies that the three zigzag orientations are no longer equivalent, and the structural anisotropy of the monoclinic structure is expected to select a preferred zigzag direction. In contrast, FePSe$_3$ maintains a threefold rotational symmetry axis in bulk form, thus requiring a multi-component magnetic order parameter to describe the zigzag magnetic ordering. This would suggest a key difference between the two compounds, with the possibility for a genuine vestigial Potts-nematic transition in FePSe$_3$. 

To investigate this possibility, we perform a combined experimental and theoretical study of rotation symmetry breaking in FePSe$_3$ using optical linear dichroism microscopy. Linear dichroism microscopy is a powerful tool to detect the breaking of threefold crystal rotation symmetry and has previously been deployed to probe zigzag magnetic order in FePS$_3$ \cite{zhangnanolett21,Xzhangnanolett21,ninanoletter2022}. We confirm the observation of signatures of vestigial nematicity, i.e., nematicity associated with the primary zigzag magnetic order, already above $T_N$. To analyze and understand these observations, we extend a theoretical model based on multicomponent Ginzburg-Landau theory \cite{Little2020} and show that the measured linear dichroism signal as a function of temperature can be explained by a vestigial order subjected to the effects of residual external strain. The need to include external strain, which explicitly breaks rotational symmetry and biases the system towards one of the nematic domains, uncovers an unexpected similarity between FePSe$_3$ and FePS$_3$, indicating that our theoretical model is applicable to both materials, and can extend to other hexagonal magnets. Whereas external symmetry breaking effects in FePSe$_3$ are likely due to the substrate or other exfoliation-induced sample inhomogeneities, in FePS$_3$ these originate from the monoclinic stacking order. While the origin of the biasing strain may be different, both materials should be understood as hosts of vestigial Potts-nematic order coupled to an external symmetry breaking field.

To set the stage for the discussion of our results, we give a brief qualitative discussion of multicomponent zigzag magnetic order and the notion of vestigial order (see also Ref. \cite{Little2020} for the case of a triangular lattice). The proper description of collinear zigzag magnetic order on the honeycomb lattice requires introducing three (real-valued) magnetic order parameters, $\bM_1$, $\bM_2$, $\bM_3$, which correspond to the three symmetry-related wave vectors $\bQ_1$, $\bQ_2$, and $\bQ_3$ of the Brillouin zone $M$ points, see Fig.~\ref{fig1}(c). In the zigzag phase only one of the three order parameters $\bM_j$ is nonzero, thus realizing single-$Q$ orders, and each ordering vector describes one of three possible orientations of the zigzag pattern related by $C_3$ symmetry.  Without magnetocrystalline anisotropy, the $\bM_j$ can point in any direction in spin space, however, significant easy-axis anisotropy in FePSe$_3$ favors ordered moments along the $z$-axis (See  Supplemental Material Section S1 for a connection between this phenomenological description and a microscopic model) . 

To characterize the discrete breaking of rotational symmetry in the single-$Q$ magnetic state, we define a subsidiary nematic order parameter $\phi$ as $\phi= \omega\bM_1^2+\omega^2 \bM_2^2+\bM_3^2$, where $\omega =e^{i2\pi/3}$. The two components of the complex nematic order parameter $( \phi_1, \phi_2) = (\text{Re}\, \phi, \text{Im}\, \phi)$ transform as the $d_{x^2-y^2}$ and $d_{xy}$ quadrupolar charge densities. In group theory language, they transform as the two-dimensional $E_g$ irreducible representation of the $\mathrm{S_6}$ point group associated with the rhombohedral space group $R \bar{3}$. Importantly, nonzero $\phi$ implies that the threefold crystallographic rotational symmetry is broken, which is clearly the case for the single-$Q$ zigzag AFM phase.  Previous theoretical studies showed that the nematic order parameter $\phi$, which is a composite of the primary magnetic order parameters $\bM_i$ can in principle order above the N\'eel temperature, giving rise to a vestigial nematic phase \cite{Fernandes2019,Little2020}. In such a phase the average staggered magnetization is zero, $\left \langle \bf{M}\right \rangle=0$, but the nematic order parameter becomes finite $\left \langle\phi\right \rangle\neq0$. Below $T_N$, in the magnetically ordered phase, both $\left \langle \bf{M}\right \rangle\neq 0$ and $\left \langle\phi\right \rangle\neq0$. This qualitative behavior is summarized in Fig.~\ref{fig1}(d).

\begin{figure}
\centering
\includegraphics[width=0.5\textwidth]{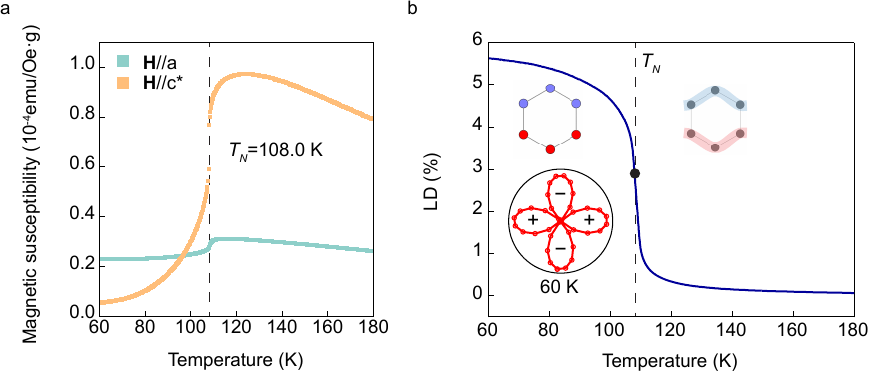}
\caption{(a) Magnetic susceptibility of the FePSe$_3$ crystal. The sharp change at 108 K indicates the formation of long-range magnetic order. (b) Temperature dependence of the linear dichroism (LD) response from a thick FePSe$_3$ flake (local point).  The black dot represents the magnetic transition temperature. Inset: The sample azimuth angle dependence of the LD response at 50 K. The `+' and `-' represent the signs of the lobes. }
\label{fig2}
\end{figure}

To study the breaking of rotational symmetry in FePSe$_3$ resulting from zigzag magnetic order, we have performed detailed optical linear dichroism (LD) measurements, which probe the reflectivity difference between different optical axes. This experimental technique has previously been used to examine the magnetic zigzag orders in FePS$_3$ \cite{zhangnanolett21,Xzhangnanolett21,zhangnatphotonic2022}. The polarization-dependent optical linear dichroism is sensitive to the breaking of $C_3$ symmetry in FePSe$_3$, and is therefore directly linked to the order parameter $\phi$. However, the linear dichroism is not directly sensitive to magnetic order and thus does not probe the N\'eel vectors; no signal is expected in magnetically ordered states that do not break rotation symmetry.  The $\textsc{LD}$ signal is defined as $\eta=\frac{R(\theta)-R(\theta+\pi/2)}{R(\theta)+R(\theta+\pi/2)}$, where $\theta$ is the angle corresponding to the light polarization and $R(\theta)$ is the angle-dependent reflectivity of the sample.  

To examine the LD signal as a function of temperature and determine whether rotation symmetry breaking occurs simultaneously with the magnetic transition, an independent accurate measurement of $T_N$ is required. We have therefore measured the magnetic susceptibility for different field directions, shown in Fig.\ref{fig2}(a), and find $T_N = 108.0\,\text{K}$. The susceptibility above $T_N$ is considerably different for fields along the $c$-axis and fields in the $ab$ plane, suggesting strong magnetocrystalline anisotropy. Furthermore, the sharp change in the susceptibility observed when the field is parallel to the $c$-axis, which becomes smaller than the in-plane susceptibility, is consistent with Ising-like spins pointing out of the plane. 

The temperature dependence of the LD from a thick FePSe$_3$ flake (around 550 nm) is plotted in Fig. \ref{fig2}(b). We observe a slow gradual increase in the LD as temperature decreases towards $T_N$, and a rapid strong rise near $T_N$. Below 105 K, the increase of the LD signal flattens. Most notably, the onset of the rapid rise in the LD already occurs above $T_N=108\,\text{K}$, which is a signature of broken rotational symmetry above the magnetic transition. Furthermore, the onset of the LD signal is smooth, in contrast to the sharp magnetic transition observed in the magnetic susceptibility. This smooth behavior is also significantly different from the behavior seen in probes of staggered N\'eel-type antiferromagnetic order realized in other phosphorous trichalcogenides, such as MnPSe$_3$~\cite{ninatnano21}. In MnPSe$_3$, the second-harmonic generation intensity, which measures inversion symmetry breaking, shows a sharp onset from zero at the magnetic transition. Importantly, the smooth onset of the LD signal observed here is qualitatively very similar to the behavior of the LD signal found in FePS$_3$~\cite{zhangnanolett21, Xzhangnanolett21,ninanoletter2022}. The difference between the two compounds is the crystal space group, as alluded to earlier. Whereas rhombohedral FePSe$_3$ has a threefold rotational symmetry, monoclinic FePS$_3$ does not. The similarity of the LD as a function of temperature is a first indicator that the underlying physics is similar, despite the different bulk structural properties. The angle dependence of the LD is shown in the inset of Fig. \ref{fig2}(b). The twofold pattern is consistent with  $C_3$ symmetry breaking.

\begin{figure*}
\centering
\includegraphics[width=1\textwidth]{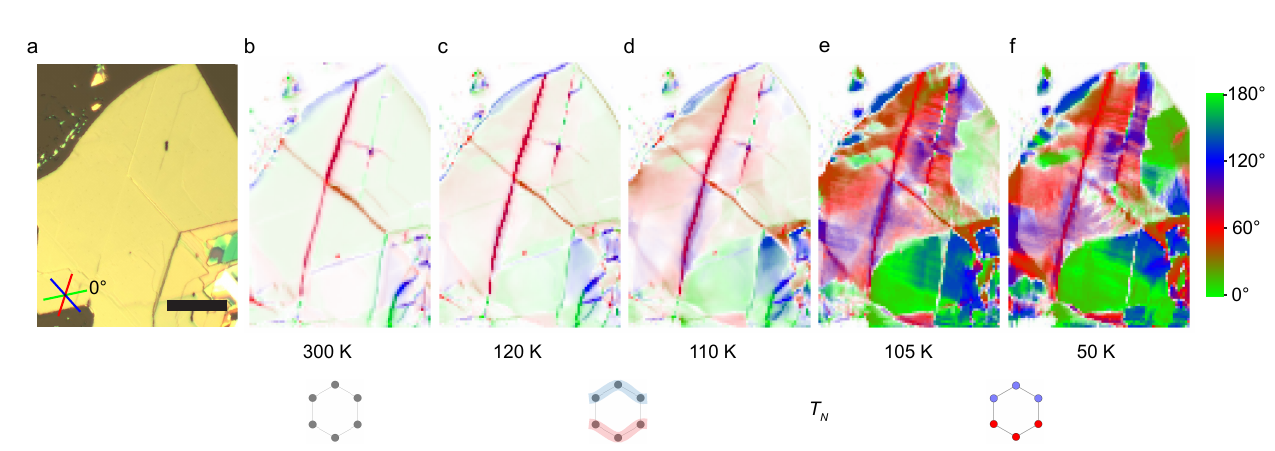}
\caption{Mapping of the order parameter $\phi$ on a thick FePSe$_3$ flake. (a) Optical image of the sample. Scale bar: 50 $\mu$m. (b)-(f) Mapping of the order parameter $\phi$ at different temperature. The color of each point represents the orientation of the order parameter $\phi$. The opacity of each point represents the magnitude of the order parameter $\phi$. The direction of the zero degree is marked in (a).}
\label{fig3}
\end{figure*}

To gain insight into the evolution of nematic order in proximity to $T_N$ and how the nematic domains develop, we conduct LD imaging on the FePSe$_3$ flake at different temperatures. The findings are illustrated in Fig. \ref{fig3}. Fig. \ref{fig3}(a) presents the optical image of the exfoliated sample situated on a SiO$_2$/Si wafer. The sample's thickness (550 nm) is determined by atomic force microscopy. Fig. \ref{fig3}(b-f) depict the spatial distribution of the LD signal across a range of temperatures. The color of each pixel (2 $\mu$m) represents the direction of the LD peak and the opacity represents the magnitude of the LD signal. These maps directly show the distribution of the direction and magnitude of the nematic order parameter $\phi$, which is what LD measures. At 300 K, the LD distribution is uniform across most of the sample, except for the artificial signals from the defects (both lines and dots visible in the optical imaging). Since a LD signal cannot be present for a crystal that has three-fold rotational symmetry, we attribute the weak signals to defects such as interlayer sliding or strains. At 120 K, the distribution of linear dichroism changes slightly and becomes non-uniform, indicating the presence of a nematic order parameter that breaks the $C_3$ symmetry. At 110 K, still above $T_N$, uniform areas break into different domains with all three possible directions, indicating the formation of vestigial $\mathbb{Z}_3$ Potts-nematic order without magnetic order. The magnitude of the anisotropy $\phi$ rapidly increases from 110 K to 105 K, accompanied by the formation of the magnetic order. We note that most of the regions have the same distribution of the anisotropy orientations between 110 K and 105 K. When we further cool the sample down to 50 K, where the zigzag antiferromagnetic order dominates, we find that the nematic order direction in some areas changes.  In our experiment, we also observe that the distribution of anisotropies at different thermal cycles remains the same in most of the sample regions.

To further analyze and interpret these observations, and in particular address the possibility of spontaneous rotational symmetry breaking above $T_N$, we now employ a Ginzburg-Landau (GL) theory for the zigzag magnetic ordering transition. The starting point of such an analysis is a GL free energy $\mathcal{F}$ expressed in terms of the magnetic order parameters $\bM_i$, which takes the form 
\begin{multline}
 \label{GL_initial}
\mathcal{F} = a \sum_i \bM_i^2 + \frac{u}{2}( \sum_i \bM_i^2 )^2 \\
	- \frac{g}{2} \left[ (\bM_1^2 + \bM_2^2 - 2 \bM_3^2)^2 + 3 (\bM_1^2 - \bM_2^2)^2 \right],
\end{multline}
Previously, this GL functional was employed to study the Potts-nematic properties of the triangular-lattice material Fe$_{1/3}$NbS$_2$ \cite{Little2020}. As is standard for GL theory, the coefficient of the quadratic term is parametrized as $a \propto T-T_c$, where $T_c$ is the mean-field magnetic transition temperature (not to be confused with the actual transition temperature $T_N$) and $g$ describes a quartic interaction between the order parameters which determines the nature of the magnetic phase below $T_N$. Here we set $g>0$ such that single-$Q$ zigzag order is favored, in accordance with experiment. To assess the onset of vestigial Potts-nematic order, we follow the large-$N$ approach previously applied to (vestigial) Ising nematic order is the context of iron-pnictides \cite{Pnictides_PRB,Fernandes2019}. Using the Hubbard-Stratonovich transformation, we decouple the quartic terms in Eq.~\ref{GL_initial} in terms of the nematic order parameter $(\phi_1,\phi_2)$, as well as a symmetric field describing Gaussian fluctuations, and integrate out the magnetic fluctuations. This yields an effective free energy $\mathcal{F}_n$ for the nematic order parameter, which can be minimized to obtain the behavior of the nematic order parameter as a function of temperature. Technical details of this derivation and analysis are presented in the Supplemental Material Section S2.

\begin{figure*}
\centering
\includegraphics[width=\textwidth]{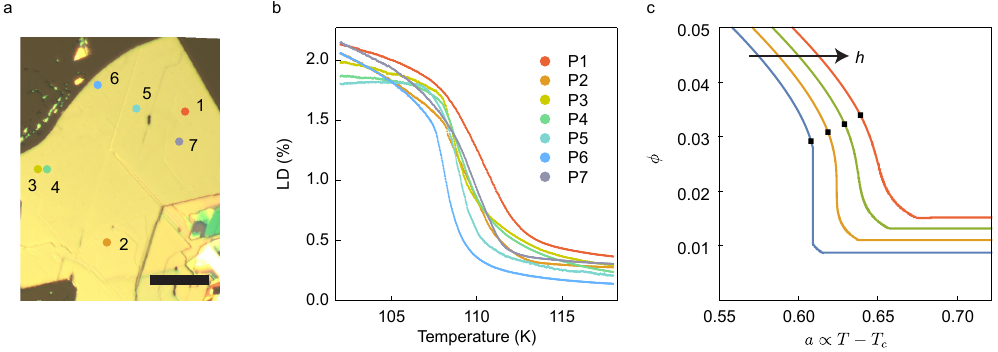}
\caption{(a)-(b) Linear dichroism signal as a function of temperature taken at different locations (i.e. micro-regions) of the sample. The position on the sample corresponding to each curve is indicated in panel (a). (c) Magnitude of the nematic order parameter $\phi$ as a function of reduced temperature $a\propto T-T_c$ calculated from the theoretical model for different values of uniform external strain $h$. We take $g/u = 0.05$ and discuss specific strain values in the Supplementary Materials Section S2.} 
\label{fig4}
\end{figure*}

The first key result of our theoretical analysis is that a vestigial nematic phase, where $ \left \langle \bf{M}\right \rangle = 0$ and $\left \langle\phi\right \rangle\neq0$, emerges when the inter-layer magnetic coupling is sufficiently smaller than the intra-layer coupling and the quartic Landau coefficient $g$ is sufficiently smaller than $u$. Outside of this parameter range, the transitions are simultaneous and first-order. In the case where the $C_3$ rotational symmetry breaking transition is split from the antiferromagnetic one, an intermediate Potts-nematic phase is realized. This $\mathbb{Z}_3$ transition, however, is first order and accompanied by a discontinuous jump of the nematic order parameter. This behavior appears inconsistent with the experimental LD measurements, which show a gradual increase of $\phi$ above $T_N$, with a sharper but still continuous increase close to $T_N$. In fact, the convex-like onset of $\phi(T)$ in Fig. \ref{fig2}(b) followed by a sharp enhancement at a well-defined temperature is typical of a first-order transition in the presence of an external field conjugate to the order parameter. In the case of nematic order this is uniaxial strain. 

We therefore extend our model by including the coupling of the nematic order parameter to a uniform external strain field $(h_1, h_2)$ in the free energy, $\mathcal{F}_n \rightarrow \mathcal{F}_n - \sum_{i=1}^2 h_i \phi_i $. To capture the shape of the LD curve, we first set the magnetic anisotropy and the ratio $g/u$ such that the system is close to the regime of simultaneous first-order $\mathbb{Z}_3$ and antiferromagnetic transitions in the absence of strain. The shapes of $\phi(T)$ for various strain values are shown in Fig.~\ref{fig4}(c), in good qualitative agreement with the experimental LD measurements of Fig. \ref{fig2}(b). Indeed, in the presence of strain, the nematic order parameter is non-zero at high temperatures and exhibits a smooth gradual increase as temperature is lowered. The absence of a sharp rotation symmetry breaking transition is expected, since the uniform strain explicitly breaks the threefold rotational symmetry and pins the nematic director.

An important question raised by this analysis is what causes the strain that is required to describe the data. Since strain is not purposefully applied on the sample, it must arise from small internal strains. This hypothesis is confirmed by measuring the LD at different locations of the sample [Fig. \ref{fig4}(a)]. As shown in Fig. \ref{fig4}(b), the nematic signals at different micro-regions not only display different onsets, but also distinct shapes. For instance, in micro-region P6, the LD curve shows a sharper enhancement than the LD curve in micro-region P1. Comparing to our theoretical results of Fig. \ref{fig4}(c), this behavior is consistent with micro-region P1 being subjected to a larger local strain than micro-region P6. 

The fact that the LD signal over the beam spot (with a diameter of 2 $\mu$m) does not average to zero implies that these small internal strains are not entirely random, but instead that the strain field consists of an average (effectively uniform) and a spatially varying random component. This becomes clearer when we compare the LD signal of FePSe$_3$ with that of FePS$_3$~\cite{Xzhangnanolett21, zhangnanolett21, ninanoletter2022}. In the latter, the monoclinic stacking of the layers creates an effective average strain field that is experienced by the layers. The remarkable resemblance between the temperature dependencies of the LD in FePSe$_3$ and FePS$_3$ indicates that the former is also subjected to an underlying strain. Since the crystal structure of FePSe$_3$ is rhombohedral, a plausible explanation for such a residual strain would be the flake's substrate.

We emphasize, however, that although the internal strains do not average to zero, they are significantly inhomogeneous. This can be seen from Fig. \ref{fig3}, which reveals that the system still breaks up in nematic domains as temperature is lowered. If there was a sufficiently large uniform strain on top of random strains, a mono-domain would be expected at low temperatures. Instead, the proliferation of domains is reminiscent of the phenomenon of domain break-up exhibited by the random-field Ising-model (RFIM) \cite{imry_random-field_1975}. To gain further qualitative insight into domain break-up in the presence of both uniform and random strain components, we numerically simulate the RFIM with a uniform field. While in the present case the random-field 3-state Potts model would be a more faithful representation of the experimental system, the computationally simpler RFIM model with uniform strain captures the essential qualitative features of domain break-up. The results, shown in Supplemental Material Section S3, reveal that  there exists a threshold uniform field strength at which domain breakup is suppressed in favor of a single domain. This is a consequence of the fact that the 2D-RFIM undergoes a transition from a disordered to an ordered phase above a threshold value of the uniform field \cite{binder_random-field_1983}. This result suggests that, in our FePSe$_3$ flakes, the residual uniform strain presumably originating from the substrate is below this critical value, which in turn is set by the distribution of internal random strains that emerge from dislocations or other lattice defects.

This analysis of strain effects, in particular the comparison with monoclinic FePS$_3$, supports our vestigial order model in which rotation symmetry breaking described by nematic order is driven by magnetic fluctuations. The uniform strain component is responsible for the gradual ``smeared'' onset of nematic order, whereas the strength of the random strain component relative to the uniform component can explain the observed domain breakup. Interestingly, in the related compound FePS$_3$, the relative strength of the two components can be controlled by the thickness of the samples, since increasing the number of layers in the monoclinic stacking sequence strengthens the uniform component. Indeed, the phenomenon of domain break-up in FePS$_3$ is only observed in thin samples and disappears as the bulk limit is approached, as expected. In contrast, domain break-up is still observed in thick near-bulk samples of FePSe$_3$. The fact that in FePSe$_3$ the magnitude of the nematic order parameter is sizable above $T_N$, despite the presumably small values of the unintentional residual strain,  provides strong support for a magnetically-driven origin of the nematic order parameter. We note that a similar observation that unintended residual strain causes a large nematic response was reported in iron-based superconductors \cite{Rosenthal2014,Baek2016}, which are the prototypical example for vestigial nematicity.

An intriguing question is whether the antiferromagnetic and the nematic transitions would occur simultaneously in the absence of strain, or would be separated by a magnetically disordered nematic phase. Our detailed theoretical analysis considering different parameter regimes, which is presented in Supplemental Material Section S2, shows that both scenarios are possible and that neither can therefore be ruled out. Looking forward, our work demonstrates that the zigzag magnetically-ordered $M$P$X_3$ magnets exhibit rich magnetic ordering phenomena due to the breaking of both crystal lattice and magnetic symmetries, thus providing a uniquely compelling venue for studying intertwined Potts-nematic, magnetic order and superconductivity in van der Waals magnets~\cite{Fernandes2019,Strockozarxiv2022}.

\section{Acknowledgments}
 The project is mainly supported by L.W.'s startup package at the University of Pennsylvania.  The development of the scanning imaging microscope
was sponsored by the Army Research Office and was accomplished under Grant Number W911NF-20-2-0166 and W911NF-21-1-0131, and the University Research Foundation.  The sample exfoliation setup is based upon work supported by the Air Force Office of Scientific Research under award number FA9550-22-1-0449. Z.N. also acknowledges support from Vagelos Institute of Energy  Science  and  Technology  graduate  Fellowship and Dissertation Completion Fellowship  at  the  University  of  Pennsylvania. D.G.M acknowledges support from the Gordon and Betty Moore Foundation's EPiQS Initiative, Grant GBMF9069. W.J.M. and R.M.F. (theoretical model and numerical calculations) were supported by the U. S. Department of Energy, Office of Science, Basic Energy Sciences, Materials Sciences and Engineering Division, under Award No. DESC0020045. W.J.M and R.M.F. thank the Minnesota Supercomputing Institute (MSI) at the University of Minnesota, where the numerical computations were performed. J.W.F.V was supported by the National Science Foundation Award No. DMR-2144352.

\section{Methods}

\subsection{Crystal growth}

Stoichiometric quantities of Fe, P, and Se powders totaling 1.5 g were mixed and sealed in a quartz tube of 11 cm length with 0.4 g I$_2$ and heated in a single zone tube furnace at 670 $^\circ$C for 7 days before being furnace cooled to room temperature. Several large crystals grew on the growth end, along with smaller (around 1mm) on both the growth and charge ends of the tube.

\subsection{Linear dichroism measurement}
The experiments are carried out using an 800 nm Ti-sapphire laser (80 MHz, 50 fs) at the normal incidence. The laser beam focused by a 50$\times$ objective is around 2 $\mu$m. The power is kept below 50 $\mu$W to minimize the laser heating. The reflected beam is collected by the same objective and measured by a photodiode. A photoelastic modulator is used to modulate the polarization at 42 kHz. A half-wave plate is used to control the polarization of the incident light. The sample position is controlled by XYZ piezo stages placed in the cryostat.\\

\textit{Data availability:} All data needed to evaluate the conclusions in the paper are present in the paper and the Supplementary Information. Additional data related to this paper could be requested from the authors.

\textit {Code availability}:  The codes used to support this study can be made available upon request.

%


\pagebreak
\widetext
\begin{center}
\textbf{\large Supplemental Materials: Observation of Z$_3$ Vestigial Potts-nematic order in van der Waals antiferromagnets}
\end{center}

\setcounter{equation}{0}
\setcounter{figure}{0}
\setcounter{table}{0}
\setcounter{page}{1}
\makeatletter
\renewcommand{\theequation}{S\arabic{equation}}
\renewcommand{\thefigure}{S\arabic{figure}}
\renewcommand{\bibnumfmt}[1]{[S#1]}
\renewcommand{\citenumfont}[1]{S#1}

\section{Magnetic order parameters of the honeycomb zigzag phase \label{OP}}

In this Supplemental section we elaborate on the definition of the three AFM ordering vectors $\bM_i$ introduced in the main text. These magnetic order parameters describe zigzag order on the honeycomb lattice and our aim in this section is to show how the order parameters can be defined starting from a microscopic spin model. This establishes a connection between the order parameters $\bM_i$ used in the Ginzburg-Landau description and the microscopic spin configuration on the honeycomb lattice. 

The precise form of the microscopic spin model is unimportant for our purpose here, which is simply to emphasize the role played by the sublattice structure of the honeycomb lattice. To illustrate the approach, consider therefore an isotropic Heisenberg model of classical spins on the honeycomb lattice of the general form
\be
H = \frac12 \sum_{ij} J_{ij} \bS_i \cdot \bS_j  = \frac12 \sum_{ij} J_{\alpha\beta}(\bq) \bS_{\bq\alpha} \cdot \bS_{-\bq\beta}. \label{eq:spin-model}
\ee
To obtain the Fourier transformed Hamiltonian we have expanded the spins in Fourier modes as
$\bS_\alpha(\bR)  = \sum_\bq \bS_{\bq\alpha} e^{i \bq\cdot (\bR + \bd_\alpha)}$, where $\alpha=A,B$ denotes the sublattice index, $\bR$ is a Bravais lattice vector, and $\bd_\alpha$ denotes the position of the $\alpha=A,B$ sites with respect to the chosen unit cell origin. Importantly, the Fourier transformed exchange couplings $J_{\alpha\beta}(\bq)$ form a matrix in sublattice space, which can be expressed as
\be
J(\bq) = \begin{bmatrix} J_{AA}(\bq) & J_{AB}(\bq)\\ J_{BA}(\bq)& J_{BB}(\bq) \end{bmatrix}. \label{J(q)}
\ee
From the perspective of a spin model given by \eqref{eq:spin-model}, the observation of zigzag magnetic order in FePSe$_3$ (and related compounds) at wave vectors $\bQ_i$ implies that the eigenvalue spectrum of $J(\bq) $, which consists of two branches, has a global minimum at $\bQ_i$. The eigenvector corresponding to the eigenvalue of the minimum at $\bQ_i$ then determines the precise structure of the (classical) magnetic ground state, i.e., the state that minimizes the classical energy. In particular, the minimal eigenvector determines how the sublattices are related. The eigenvectors can be distinguished and labeled by the symmetries which leave $\bQ_i$ invariant. One of those symmetries is inversion symmetry, which gives rise to even and odd solutions under inversion. Since inversion symmetry exchanges the honeycomb sublattices, the sublattice structure of the magnetic ground state is fully determined by its symmetry properties under inversion. In the specific case of the zigzag ordered state, the corresponding eigenmode is odd under inversion and the zigzag phase therefore can be called sublattice-odd. (The other solution at $\bQ_i$ is sublattice-even and corresponds to the so-called stripe phase.) More precisely, the zigzag eigenmode solution can be written as
\be
\bS(\bQ_i) = v_A \bS_{\bQ_i A} + v_B \bS_{\bQ_i B}, \label{S(Q)-definition}
\ee
where $(v_A, v_B)$ is the corresponding eigenvector of $J(\bQ_i)$. The magnetic order parameters $\bM_i$ are then defined as $\bM_i =\langle \bS(\bQ_i) \rangle$. The spin configuration in real space can be expressed in terms of $\bS(\bQ_i) $ by appropriately inverting \eqref{S(Q)-definition}. 

For the purpose of this brief discussion, we have used an isotropic model given by \eqref{eq:spin-model}, which does not accurately capture the magnetocrystalline anisotropies observed in FePSe$_3$ and other materials. It is, however, sufficient to illustrate the general method of analysis, and in particular to show that zigzag order on the honeycomb lattice is described by three symmetry-related magnetic order parameters. It furthermore exposes where information about the honeycomb sublattices is hiding. When including anisotropies, which would simply promote \eqref{J(q)} to a larger matrix in sublattice and spin space, the structure of the analysis remains the same. In particular, the way in which the sublattices are related---by the symmetry properties of the eigenstates of $J(\bq)$---remains the same. Magnetocrystalline anisotropy can be included by adding a single-ion anisotropy term $H_\Delta = \Delta \sum_i (S^z_i)^2$ to \eqref{eq:spin-model}. Since experiments show easy-axis anisotropy in FePSe$_3$, this suggests $\Delta<0$ and implies that the magnetic order parameters $\bM_i$ take the Ising-like form $M^z_i$.

\section{Ginzburg-Landau theory of Potts-nematic vestigial order \label{GL}}

As described in the main text, the starting point of our theoretical model is a Ginzburg-Landau (GL) expansion of the free energy in terms of the magnetic order parameters, i.e., 
\be \label{GL_initial}
	\bar{\mathcal{F}} = \sum_i \left( a  \bM_i^2 + (\nabla \bM_i)^2 \right) + \frac{u}{2} \left( \sum_i \bM_i^2 \right)^2 - \frac{g}{2} \left[ (\bM_1^2 + \bM_2^2 - 2 \bM_3^2)^2 + 3 (\bM_1^2 - \bM_2^2)^2 \right],
\ee
where $\bM_i  = \bS (\bQ_i)$ are the Fourier components of the spin density with wavevectors $\bQ_i$, see Sec.~\ref{OP}.
Gradient term $(\nabla \bM_i)^2$ was added as we will account for the spatial fluctuations of magnetization (we choose the units that it has no prefactor).
As discussed in the main text, this phenomenological model has been previously scrutinized in Refs. \cite{Fernandes2019,Little2020}. Since the wave-vectors are commensurate and correspond to the three $M$ points of the Brillouin zone, one has $\bQ_i \sim -\bQ_i$, which implies that the order parameters are real. We assume that the coefficient $u$ is positive to ensure that the free energy is bounded from below (strict stability criterion is $u > 4g$). 

The structure of the magnetically ordered state is determined by the minimum of \eqref{GL_initial} at $a < 0$. Here we choose the fourth order coefficient $g>0$ such that the magnetic state is characterized by single-$Q$ ordering, i.e., only one out of three $\bM_i$ is nonzero. Therefore, the magnetic state not only breaks the spin rotational symmetry, as well as time-reversal symmetry, but also the $C_3$ rotational symmetry of the lattice. 
To study the possibility of a rotational symmetry breaking transition preempting the magnetic transition, we introduce the Potts-nematic order parameter $(\phi_1, \phi_2)$ defined in terms of the magnetic order parameters in the main text, and perform a Hubbard-Stratonovich (HS) decoupling of the free energy, following the same steps as \cite{Fernandes2019,Little2020}. The HS decoupling yields a free energy given by
\be 	\label{GL_decoupled}
	\bar{\mathcal{F}}[r, \phi_i,\bM_i] = \sum_i \left(r  \bM_i^2 + (\nabla \bM_i)^2 \right) - \frac{(r - a)^2}{2u} + \frac{\phi^2}{2g} -\phi_1 (\bM_1^2 + \bM_2^2 - 2 \bM_3^2) - \sqrt{3} \phi_2 (\bM_1^2 - \bM_2^2),
\ee
where we also introduced an auxiliary field $r$, which describes symmetric Gaussian fluctuations of the magnetic order.  The original GL free energy \eqref{GL_initial} is recovered from \eqref{GL_decoupled} by performing Gaussian integration over the fields $r$ and $\phi_i$. 

Since the free energy \eqref{GL_decoupled} is now quadratic in the magnetic variables $\bM_i$, these can be integrated out using standard Gaussian integration. When performing the integration we follow the approach of Ref.~\onlinecite{Pnictides_PRB} and treat the magnetic variables $\bM_i$ as (classical) vectors with $N$ components. This step is taken to make the theory formally controlled in the limit where $N$ is large, and a large-$N$ expansion can thus be pursued. In particular, in the large-$N$ limit it is formally justified to perform a saddle-point analysis of the resulting free energy. 

Note that since $N$ denotes the components of spin, $N=3$ corresponds to a Heisenberg magnet, $N=2$ corresponds to an XY magnet, and $N=1$ corresponds to an Ising magnet. The analysis of the magnetic susceptibility observed in FePSe$_3$ suggests appreciable Ising anisotropy, with moments along the $z$ axis perpendicular to the honeycomb planes. One may thus reasonably expect that FePSe$_3$ can be described by an effective dimensionality $1 < N < 3$, although formally Ising anisotropy leads to an Ising fixed point. As will be clear from the remainder of the calculation, the theory derived in the limit of large $N$ does not bear any essential dependence on $N$, which can be  eliminated by redefining phenomenological coefficients of the Ginzburg-Landau theory. 

We proceed to performing the integration over $\bM_i$, which yields a free energy that is a function of the $r$ and $\phi$ variables only, and reads as (see also Ref. \cite{Little2020})
\be \label{F_fermion_integrated}
	\bar{\mathcal{F}}[r, \phi_i] = \frac{N}{2} \int d^d\bq \, \tr \log \hat{\chi} (\bq) - \frac{(r - a)^2}{2u} + \frac{\phi^2}{2g}. 
\ee
Here, $\hat\chi$ is a matrix of spin susceptibilities given by
\be
\hat{\chi}^{-1} =(r + q^2)I_{3} - 2\phi\begin{pmatrix}\cos (2\theta - \kappa_1)&0&0 \\ 0&\cos(2\theta - \kappa_1)&0 \\ 0&0&\cos (2\theta -  \kappa_3) \end{pmatrix},
\ee
with $(\kappa_1,\kappa_2,\kappa_3) = (\pi/3,5\pi/3,\pi)$, and we have furthermore introduced the notation $(\phi_1, \phi_2) = \phi (\cos 2\theta, \sin 2 \theta)$. The angle $\theta$ may then be interpreted as the direction of the nematic director. In the case of one single layer, the momentum integral in \eqref{F_fermion_integrated} is two-dimensional ($d=2$), which prohibits magnetic ordering due to the Mermin-Wagner theorem. 
In reality, the system under consideration has a three-dimensional structure resulting from the stacking of individual layers. In Ref.~\onlinecite{Pnictides_PRB} it was shown that, to account for the coupling in the third dimension, one can may take $d$ to be an effective fractional dimension with $2 < d < 3$. We will apply the same method in our derivation.

We take the momentum integral in \eqref{F_fermion_integrated} imposing a hard cutoff $q^2 < \Lambda$ at large momenta. Then, the free energy takes the form (up to an irrelevant constant term and overall rescaling):
\be \label{F_integrated}
	\mathcal{F}[r, \phi_i] =  -\frac{(r - \bar{a})^2}{2 \bar{u}} + \frac{\phi^2}{2\bar{g}} - \frac{N}{d} \sum_i \left[ r - 2\phi \cos (2\theta - \theta_i) \right]^{d/2},
\ee
where $\theta_1 = \pi/3$, $\theta_2 = 5\pi/3$, and $\theta_3 = \pi$ and the renormalized coupling constants were redefined as $(\bar{u}, \bar{g}) = (u, g) \cdot \Gamma(2-d/2) / (2^{d-1} \pi^{d/2} (d-2))$, and $\bar{a} = a + 3 \bar{u} \Lambda^{d-2} / (2 \Gamma(2 - d/2) \Gamma(d/2) )$. The renormalization of $a$ is due to nonuniversal contributions of high momenta and can be interpreted as an irrelevant shift of the critical temperature. 

We then proceed within the framework of a large-$N$ expansion \cite{Pnictides_PRB} and apply the saddle-point approximation analysis to the free energy \eqref{F_integrated}, i.e., we minimize it with respect to $r$, $\phi$, and $\theta$ parameters.
It follows from \eqref{F_integrated} that the most energetically favorable directions of the Potts-nematic director correspond to the three values $\theta = \pi/6, \pi/2, 5\pi/6$. 
So as a first step of the minimization, we pick a specific value $\theta = \pi/2$, which means that $\phi_1 = -\phi, \phi_2 = 0$. Note that any further minimization is subject to certain constraints. 
Indeed, as is evident from the expression of $\eqref{F_fermion_integrated}$ and the discussion that follows it, the magnetic susceptibility diverges at $r = 2\phi$. 
Since the derivation above was made under the assumption of zero magnetization, one should therefore restrict the minimization domain to the region $0 < \phi < r/2$. (Note $\phi>0$ since $\phi$ is the magnitude of the nematic order parameter, and thus non-negative by definition.) 

\begin{figure}[h]
	\begin{center}
	\includegraphics[width=0.5\textwidth]{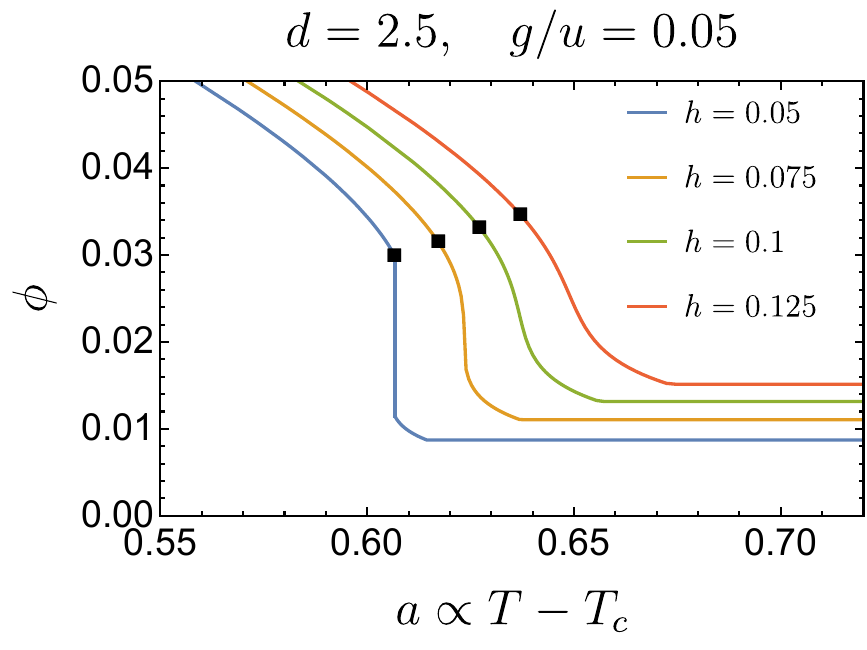}
	\caption{Potts-nematic order parameter as a function of temperature for a number of applied strain values $h$ with the nematic director aligned along the strain direction. Black squares mark the onset of the magnetic order, which can happen at lower temperatures compared to the uprise of nematic order parameter. We introduced the effective dimension $d=2.5$ to account for the layered structure of the material, which suppresses strong two-dimensional fluctuation effects. Here $\bar{u} = 1$, $\bar{g} = 0.05$.}
	\label{fig:dim25}
\end{center}
\end{figure}

To account for the influence of strain on the system properties, we note that it couples linearly to the Potts-nematic order parameter and thus can be introduced to the GL theory in the following way:
\be
	\mathcal{F}_s = \mathcal{F} - \sum_i h_i \phi_i .
	\label{GL_decoupled_strain}
\ee
Arbitrary strain can deflect the direction of the Potts-nematic order parameter from the three possible values that are favorable in the unstrained system. 
However, for simplicity below we will assume that strain promotes the Potts-nematicity in the picked direction of $\theta = \pi/2$, which corresponds to $h_1 = h$ and $h_2 = 0$.
Then, the action reads:
\be \label{F_for_minimization}
	f_s = -\frac{(r - \bar{a})^2}{2 \bar{u}} + \frac{\phi^2}{2\bar{g}} - \frac{1}{d} \left( 2 (r + \phi)^{d/2} + (r - 2\phi)^{d/2} \right) - h \phi ,
\ee
where we also redefined the GL parameters $(\bar{u},\bar{g}) \rightarrow (\bar{u}, \bar{g}) / N $ and factored out $N$ from the free energy: $f_s = \mathcal{F}_s / N$. 
As mentioned above, this procedure completely eliminates $N$  from the equations. 
In the process of minimization, it is convenient to use the variable $\varphi = 2 \phi /r$ instead of $\phi$, because the former takes values in the interval $0 < \varphi < 1$ in the magnetically disordered phase. 
We proceed by numerically minimizing \eqref{F_for_minimization} with respect to $r$ and $\varphi$. Technically, we express $r(\varphi)$ as a function representing the numerical solution of the equation $\partial f_s / \partial \varphi = 0$ and then minimize the free energy with respect to $\varphi$, decreasing $a$ step-by-step and using the position of the previous minimum as an initial guess for the next value of $a$. 
The results of this procedure at $\bar{u} = 1$, $\bar{g} = 0.05$, effective dimension $d = 2.5$ and a number of different strain values are presented in Fig.~\ref{fig:dim25}, corresponding to Fig. 4(c) of the main text, where we plot the absolute value of the Potts-nematic order parameter $\phi$ against the GL parameter $a$ in arbitrary units. Since near the critical temperature $T_c$ (and critical $a = a_c$) the linear expansion $a - a_c \propto T - T_c$ is valid, one can interpret Fig.~\ref{fig:dim25} as the plot of the $\phi(T)$ dependence. 
The figure demonstrates that when the temperature is lowered, the Potts-nematic order parameter emerges following a peculiar curve, whose shape depends on the value of the strain. The strain also leads to a nonzero value of $\phi$ at large temperatures, since it acts as a symmetry-breaking field. At a certain point, the minimum of the free energy \eqref{F_for_minimization} on the interval $0 < \varphi <1$ reaches the point $\varphi = 1$, which corresponds to $r = 2\phi$, the point where the magnetic susceptibility diverges [see $\eqref{F_fermion_integrated}$ and the the paragraph below]. That marks the magnetic phase transition, which is shown in Fig.~\ref{fig:dim25} by black squares. 

\begin{figure}[h]
	\begin{center}
	\includegraphics[width=0.5\textwidth]{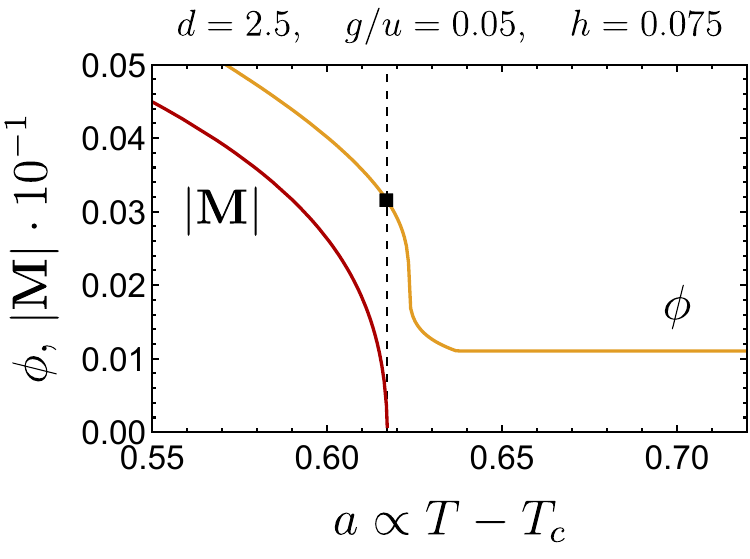}
	\caption{Potts-nematic and magnetic order parameters as a function of temperature in the effective dimension $d=2.5$ for a selected strain value $h = 7.5 \cdot 10^{-2}$ and $\bar{u} = 1$, $\bar{g} = 0.05$. The uprise of $\phi$ preempts the magnetic phase transition. }
	\label{fig:phi_and_M}
\end{center}
\end{figure}

To extend the $\phi(T)$ dependence below the magnetic transition, one has to modify the theory presented above. For that one should repeat the derivation including nonzero average value for one of the $\bm{M}$ components. For our choice $\theta = \pi/2$, the diverging susceptibility corresponds to the $\bm{M}_3$ vector, so we consider $\bm{M}_3  = M \bm{n} + \tilde{\bm{M}}_3$, where $\bm{n}$ is an arbitrary direction in the spin space and in the following we will suppress tilde in $\tilde{\bm{M}}_3$. The derivation leads to the following modification of the action:
\be \label{F_with_M}
	f_M [M, \phi, r] = \mathcal{F} / N + M(r - 2\phi),
\ee
valid in the large-$N$ limit, where we also rescaled $M^2 \rightarrow M^2 N$.
This free energy should be extremized with respect to $M$, $r$, and $\phi$ variables. The condition $\partial f_M / \partial M = 0$ leads to the constraint $r = 2\phi$, so that $M$ acts as a Lagrange multiplier. Then, the free energy is reduced to the form
\be \label{F_at_varphi1}
	f'_M = -\frac{(2\phi - \bar{a})^2}{2 \bar{u}} + \frac{\phi^2}{2\bar{g}} - \frac{2}{d}  (3 \phi)^{d/2} - h \phi .
\ee
We minimize \eqref{F_at_varphi1} to obtain the parts of the curves in Fig.~\ref{fig:dim25} located to the left of the black squares, which mark the magnetic transition. 
One can also obtain the magnetization $M$ as a function of temperature by substituting the obtained values of $r$ and $\phi$ to the equation $\partial f_M / \partial \phi = 0$ and solving it with respect to $M$.
We present both Potts-nematic and magnetic order parameters in Fig.~\ref{fig:phi_and_M} for a selected value of strain.

\begin{figure}[h]
	\begin{center}
	\includegraphics[width=0.95\textwidth]{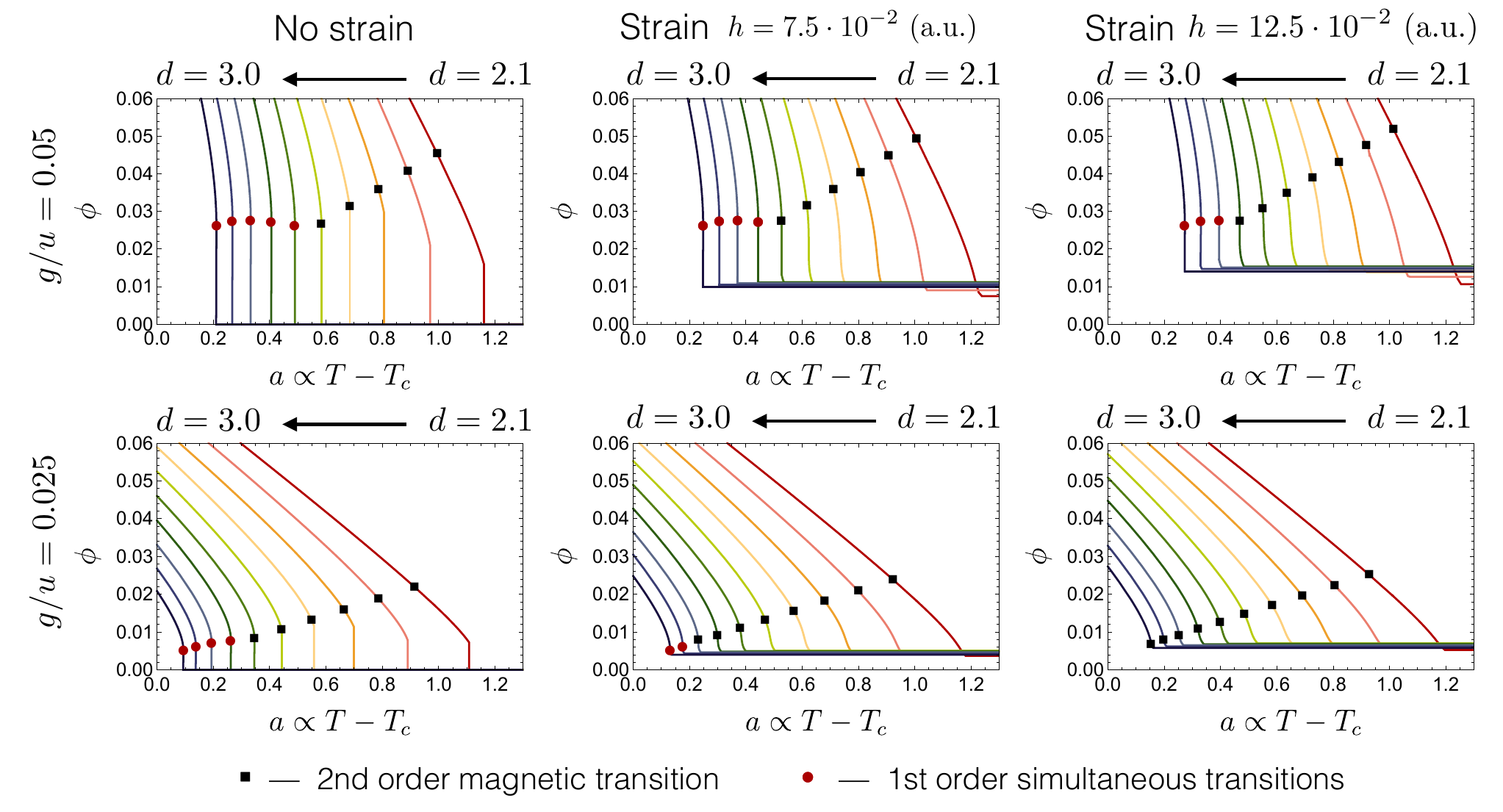}
	\caption{Dependence of the nematic order parameter on the GL parameter $a \propto T - T_c$ in different effective dimensions $d = 2.1 \div 3.0$. The strain is absent (left column), equals $h = 0.075$ (central column), or $h = 0.125$ (right column). The ratio of the GL parameters is $g/u = 0.05$ (upper row) and $g/u = 0.025$ (lower row). Magnetic transition is shown as black square (second order) or red dot (first order, simultaneous with the jump in $\phi$).}
	\label{fig:different_dim}
\end{center}
\end{figure}

To further study the interplay of  fluctuations, strain, and interlayer coupling, we perform the aforementioned analysis in different effective dimensions $2.1 \leq d \leq 3.0$ (Fig.~\ref{fig:different_dim}). We consider the case without strain and with $h = 0.075, \, 0.125$. The calculation is done for $\bar{u} = 1$ and for two values $\bar{g} = 0.05, \, 0.025$.
For the studied parameter values, we observe that the magnetic phase transition can be of second order (shown in black squares, compare with Fig.~\ref{fig:phi_and_M}) and thus split from the nematic transition or first order (red dots). In the latter case it is accompanied by a jump in the nematic order parameter $\phi$. The nematic order parameter itself is generally nonzero even at large temperatures in the presence of the symmetry-breaking strain field. Note that the relative horizontal shift between the curves inside one picture is not physical due to the ultraviolet contribution to $a$ that renormalizes $T_c$.

Fig.\,\ref{fig:different_dim} demonstrates that as the effective dimension is lowered (which corresponds to a weaker interlayer coupling) the width of the vestigial phase hosting enhanced values of $\phi$ is broadened while the magnetic transition becomes second-order. It is natural as eventually at $d=2$ there is no magnetism due to the Mermin-Wagner theorem. A similar effect of a broadened vestigial phase regime is observed upon decreasing $g/u$. By comparing the curves for different values of $d$ or $g/u$ in the presence of strain, we note that several of them share similar qualitative features regardless of the whether the magnetic and nematic transitions for zero strain are split or simultaneous.

\section{Suppression of domain-breakup by external strain}

In this section, we detail our numerical simulations used to understand the 3-state Potts nematic domain breakup seen in the FePSe$_3$ flakes. While a simulation of the random 3-state Potts model would provide a closer analogy with the experiment, the essential physics is captured instead by the conceptually and computationally simpler random-field Ising model (RFIM), defined by 
\be
H = -J\sum_{\left\langle ij\right\rangle} S^z_i S^z_j - \sum_i h_i S^z_i,
\ee
where $J \equiv 1$ sets the energy scale of the problem and the notation $\left\langle ij \right\rangle$ denotes a sum over nearest-neighbor bonds. In this case, the Ising pseudospins with values $S_i^z = \pm 1$ corresponds to the local Ising nematic order parameter. The site-dependent fields $h_i$ are spatially uncorrelated with mean $h$ and variance $\delta$. These random fields correspond to random strains that locally break the symmetry of the nematic order parameter.

It is known that, in 2D, the ordered state of the Ising model destabilizes in the presence of a random field with $h= 0$, even at zero temperature \cite{imry_random-field_1975, binder_random-field_1983, AizemannUniqueGSRFIM}. Indeed, the ground state breaks apart into Ising domains at a minimum length scale $\ell_b \sim \exp\left(\mathcal{C} J^2/\delta^2\right)$, with $\mathcal{C}$ being a constant of order one \cite{binder_random-field_1983}. While these calculations were performed for the Ising model, with $Z_2$ symmetry, the resulting domain breakup in 2D is expected for other random-field systems with a discrete symmetry \cite{Aharony1984}. In FePSe$_3$, for example, the nematic order parameter has a $Z_3$ discrete symmetry, and one expects domain breakup at low temperatures when it experiences local random strains generated by structural disorder.

As detailed in Ref. \cite{binder_random-field_1983}, the random field physics is richer when the random field distribution is biased with a mean $h\neq 0$. This case corresponds simply to random fields on top of a uniform external field. For a nonzero bias, the ordered state is restored by driving the system through a first-order transition with a critical external field $h_c$ given by 
\be
h_c = \frac{4\delta^2}{\mathcal{C} J} \exp\left [-1 - \mathcal{C}\left( \frac{J}{\delta} \right)^2 \right].
\ee
For FePSe$_3$, this critical strain would be parametrically small in the structural disorder strength $\delta$. Because of the first-order nature of this transition, close to $h_c$ one expects phase coexistence, but in this situation it would be between a large uniform nematic region, and other regions riddled with domains. Therefore, upon cooling, one would expect to see both phenomena exhibited by the system, with the final ground state appearing only after freezing out its counterpart metastable state. The ground state would present as a single uniform domain, or many smaller intermixed domains, depending on whether $h > h_c$ or $h < h_c$, respectively.

To simulate the domain breakup seen in FePSe$_3$ in Figure 3 of the main text, we apply a simulated annealing protocol to a $512 \times 512$ RFIM on a periodic square lattice comprised of $262\, 144$ Ising pseudospins. Our simulated annealing protocol is ``adaptive" in the sense that the temperature is lowered only if the simulation detects that thermal equilibrium has been reached by comparing successive time-averages of the energy and specific heat. Details of this algorithm can be found in Appendix B of Ref. \cite{meese_random_2022}, while those for our code can be found in \footnote{Our simulations were written in the \textit{Julia} programming language \cite{bezanson2017julia} and made heavy use of the \textit{JLD2} package for data compression, \textit{DrWatson} package for data and directory management \cite{DrWatson2020}, and \textit{Makie} package for plotting \cite{makiecitation}. Our simulations were performed at the \textit{Minnesota Supercomputing Institute} on the \textit{Agate} cluster.} and references therein.

In our simulations, $J = 1$, and we used Gaussian random fields with fixed $\delta = 0.95$. Such a large value was chosen to generate many domains within the lattice at zero bias $h=0$. Each simulated annealing regimen started from a random pseudospin configuration at a high temperature of $T_H = 8$ and was cooled to a low temperature of $T_L = 0.1$ over 50 temperatures. The relative temperature change was fixed to be about 8.9\%, and, at each temperature, $40\, 960$ Metropolis \cite{metropolis_equation_1953, gubernatis_quantum_2016} sweeps were performed in each adaptive equilibration block \cite{meese_random_2022}.

\begin{figure}
    \centering
    \includegraphics[width=\columnwidth, keepaspectratio]{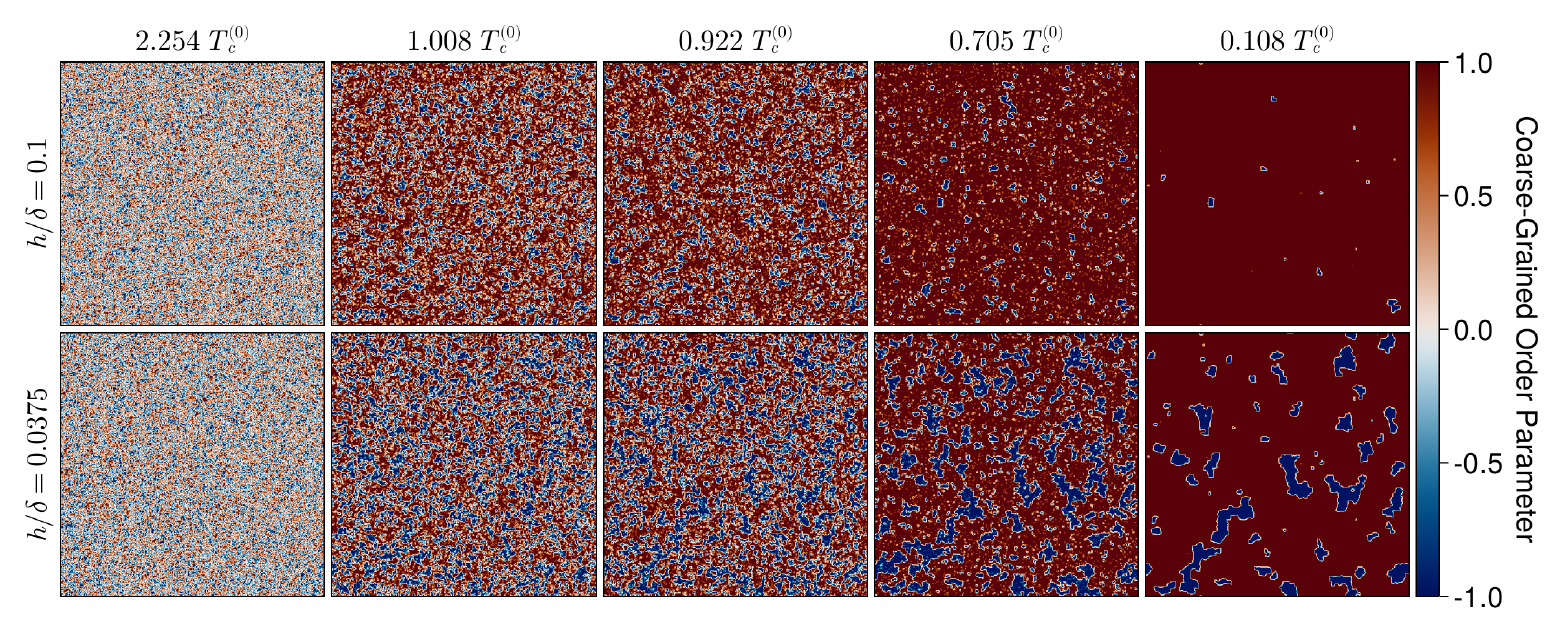}
    \caption{Snapshots of the coarse-grained order parameter of a RFIM undergoing adaptive simulated annealing. The rows correspond to the same system with a bias $h / \delta$, whereas the columns show the temperature $T$ in units of the clean 2D Ising model's transition temperature $T_c^{(0)} = 2J/\mathrm{arcsinh}(1) \approx 2.269\, J$. Both rows show a bias towards a positive order parameter upon cooling, but the system with the lower bias breaks apart into more domains at the lowest temperatures.}
    \label{fig:sa_protocols}
\end{figure}

We increased the bias of the random field distribution from $h = 0$ and found that at $h/\delta \approx 0.038$, the ground state domains contracted significantly. For larger $h$ values, the domains nearly vanished. In Figure \ref{fig:sa_protocols} we compare snapshots of two adaptive simulated annealing protocols, one with $h/\delta = 0.1$ and the other just below the critical bias at $h/\delta = 0.0375$. We plot the field of coarse-grained Ising pseudospins to simulate the local nematic order parameter $\phi_i$ defined by 
\be
\phi_i = \frac{1}{8} \sum_{j \in \mathcal{V}(i)} S_j^z.
\ee
In the expression above, $\mathcal{V}(i)$ is the set of eight nearest and next-nearest neighbors of site $i$. 

Both systems were initialized from the same random pseudospin and field configuration; thus, the only difference between the two protocols is the effect of the external field $h$ on the Metropolis acceptance probabilities. One can see that both systems tend to favor $\phi > 0$ due to the their positive biases $h$ as they cool from high temperatures. However, at low temperatures, only the system with the lower $h$ value exhibits large-scale domains with $\phi < 0$, whereas the system with the higher $h$ bias tends to flip most domains and form a nearly uniform $\phi = 1$ state. 

The tendency shown in the bottom panels of Fig. \ref{fig:sa_protocols} of domain breakup even in the presence of an external conjugate field is qualitatively similar to the domain behavior seen via optical linear dichroism in FePSe$_3$ as shown in Figure 3 of the main text. The main difference, of course, is that experimentally there are three nematic states illustrated with green, red, and blue Potts domains, whereas in our simulations there are two Ising-nematic domains -- red regions with $\phi > 0$ and blue regions with $\phi < 0$.  Focusing on the upper-left region of the sample shown in Figure 3 of the main text, we note that the optical linear dichroism measurements, particularly between 120 and 110 K, indicate that the strain in that region favors the red nematic domain, with green and blue being disfavored. It is only upon cooling that the disfavored green and blue nematic domains start to appear in this region of the sample and then at the lowest temperatures stand out against the red background. This behavior is reminiscent of the domain breakup in our simulated RFIM that occurs at low temperatures despite the presence of the external strain field. 

\end{document}